\documentclass{aa} 

\usepackage{graphicx}
\usepackage{txfonts}
\usepackage{natbib}
\usepackage{url}
\bibpunct{(}{)}{;}{a}{}{,}

\newcommand{\be}{\begin{equation}}
\newcommand{\ee}{\end{equation}}

\def\apjl{ApJL}
\def\apjs{ApJS}

\newcommand{\Mpc}{$h^{-1}$\thinspace Mpc}

\newcommand{\vmh}{h^{-1}\mathrm{Mpc} }

\begin{document}  

\title{SDSS superclusters: morphology and galaxy content
} 

\author {M.~Einasto\inst{1} \and H. Lietzen \inst{1,2,3} 
\and E.~Tempel\inst{1,4} 
\and M.~Gramann\inst{1} 
\and L.J.~Liivam\"agi\inst{1,5} 
\and J.~Einasto\inst{1,6,7}  
}

\institute{Tartu Observatory, 61602 T\~oravere, Estonia
\and
Instituto de Astrofísica de Canarias, E-38200 La Laguna, Tenerife, Spain
\and
Universidad de La Laguna, Dept. Astrofísica, E-38206 La Laguna,  
Tenerife, Spain
\and 
National Institute of Chemical Physics and Biophysics, R\"avala pst 10,
Tallinn 10143, Estonia
\and
Institute of Physics, Tartu University, T\"ahe 4, 51010 Tartu, Estonia
\and
Estonian Academy of Sciences,  EE-10130 Tallinn, Estonia
\and 
ICRANet, Piazza della Repubblica 10, 65122 Pescara, Italy
}

\authorrunning{M. Einasto et al. }

\offprints{M. Einasto}

\date{ Received   / Accepted   }

\titlerunning{Sclgal}

\abstract
{
Understanding the formation, evolution and present-day properties
of the cosmic web and objects forming it is an important task in cosmology. 
}
{
We compare the galaxy populations in superclusters of different morphology
in the nearby Universe (180~\Mpc\ $\le d \le $ 270~\Mpc)
to see whether the inner structure and overall morphology
of superclusters are important in shaping 
galaxy properties in superclusters.
}
{
We find supercluster morphology with Minkowski functionals and 
analyse the probability density distributions of colours,
morphological types, stellar masses, star formation rate (SFR) of galaxies,
and the peculiar velocities of the main galaxies in groups
in superclusters of filament and spider types, and in the field.
We test the statistical significance of the results with the 
KS test.
}
{
The fraction of red, early-type, low SFR 
galaxies in filament-type superclusters 
is higher than in spider-type superclusters;
in low-density global environments their fraction is lower than in superclusters. 
In all environments the fraction of red, high stellar mass, 
and low SFR galaxies in rich  groups 
is higher than in poor groups.
In superclusters of spider morphology red, high SFR galaxies 
have higher stellar masses than in filament-type superclusters.
Groups of equal richness host galaxies with larger stellar masses, 
a larger fraction of early-type and red galaxies, 
and a higher fraction of low SFR galaxies, if they are located 
in superclusters of filament morphology.
The peculiar velocities of the main galaxies in groups from
superclusters of filament morphology are higher than in those of
spider morphology. Groups with higher peculiar velocities of 
their main galaxies in filament-type superclusters are located in higher
density environment than those with low peculiar velocities.
 There are significant differences between galaxy populations 
 of the individual richest superclusters.
}
{ 
Both local (group) and global (supercluster) environments and 
even supercluster morphology play an important role in the 
formation and evolution of galaxies. Differences in the inner structure
of superclusters of filament and spider morphology and the dynamical state 
of galaxy groups in them may lead to the differences found in our study.
}


\maketitle

\section{Introduction} 
\label{sect:intro} 

The large scale structure of the Universe is formed by a hierarchy of galaxy systems
from isolated galaxies to groups, clusters, and superclusters of galaxies.
Galaxy superclusters, environments for galaxies, and groups and clusters of galaxies,
are the largest relatively isolated systems 
in the Universe are galaxy superclusters \citep{1956VA......2.1584D, 1978MNRAS.185..357J, 1984MNRAS.206..529E,
1994MNRAS.269..301E, 1993ApJ...407..470Z}. 
To understand the role of small-scale (group) and large-scale (supercluster) 
environments in galaxy formation and evolution, we need to study 
the properties of superclusters and their galaxy and group populations together.

With the advent of deep surveys covering
large areas in the sky it became possible to determine the parameters of 
a large number of superclusters, and to study in detail
the galaxy and group populations in them.
These studies have revealed that in richer superclusters
groups and clusters of galaxies are also richer, and they
contain a larger fraction of red, non-star forming galaxies than
poor superclusters \citep{2007A&A...462..397E, 2007A&A...464..815E}.

Early studies of galaxy environment showed that environments at the group
and cluster scale affects the morphology of galaxies, early-type, 
more luminous galaxies are located in higher density environments
than late-type, fainter galaxies \citep{1974Natur.252..111E, 1980ApJ...236..351D, 
1984ApJ...281...95P, 1987MNRAS.226..543E, 1988ApJ...331L..59H, 1991MNRAS.252..261E}.
The environment affects the properties of
galaxies and groups of galaxies 
even at larger  scales, up to 10--15~\Mpc\
\citep{1987MNRAS.226..543E, 1992MNRAS.255..382M, 2003A&A...401..851E, 
2003A&A...405..821E, 2009MNRAS.392.1467S, 2011A&A...529A..53T}.
In addition, groups of the same richness
contain a larger fraction of red, non-star-forming  galaxies if they are 
located in rich superclusters compared to groups in poor
superclusters or in low-density large-scale environments outside
of superclusters \citep{2008ApJ...685...83E, 2012A&A...545A.104L}.
\citet{2013MNRAS.432.1367L} showed that the stellar population of galaxies 
in groups in superstructures is systematically older,
and groups themselves are larger and have
higher velocity dispersions than groups which do not belong to
superstructures.

The morphology of superclusters have been studied, for example,  by
\citet{2011MNRAS.411.1716C} and \citet{2007A&A...476..697E}. 
Galaxy superclusters have complex inner structures 
that can be quantified with morphological
descriptors as Minkowski functionals. \citet{2007A&A...476..697E,
2011A&A...532A...5E} found in the wide 
morphological variety of superclusters two main types of superclusters:
filaments and spiders. In filaments high-density core(s) of superclusters
are connected by a small number of galaxy chains (filaments).  In spiders there 
are many galaxy chains between high-density cores in
superclusters. Poor spider-type superclusters are similar to our Local
supercluster with one rich cluster and filaments of galaxies and poor groups of
galaxies around it \citep{2007A&A...476..697E}.  

\citet{2012A&A...542A..36E} 
analysed the structure of rich galaxy clusters in superclusters
of different morphologies and showed that clusters in superclusters of spider morphology 
have higher probabilities of having substructure
and their main galaxies have higher peculiar velocities than clusters 
in superclusters of filament morphology.  

In the present study we continue our analysis of superclusters
of different morphology, studying their galaxy and group populations, to understand
better how the large-scale (supercluster) environment affects
the properties of galaxies and groups residing in them.  
For comparison we also analyse galaxy content and group
properties in low-density global 
environment (field).

In Sect.~\ref{sect:data} we describe the 
data about galaxies, groups, and superclusters used in this paper, 
in Sect.~\ref{sect:results} we compare the galaxy content and the
peculiar velocities of the main galaxies in groups in superclusters
of different type, and in the field. We discuss the 
results and draw conclusions in Sect.~\ref{sect:discussion}.

We assume  the standard cosmological parameters: the Hubble parameter $H_0=100~ 
h$ km~s$^{-1}$ Mpc$^{-1}$, the matter density $\Omega_{\rm m} = 0.27$, and the 
dark energy density $\Omega_{\Lambda} = 0.73$ \citep{2011ApJS..192...18K}.

\section{Data} 
\label{sect:data} 

\subsection{Galaxy, group, and supercluster data}
\label{subsec:dat}

We use the MAIN sample of the 8th data release of the Sloan Digital Sky 
Survey (SDSS DR8) \citep{2011ApJS..193...29A} with the 
apparent Galactic extinction corrected $r$ magnitudes $r \leq 
17.77$, and the redshifts $0.009 \leq z \leq 0.200$, in total 576493 galaxies. 
We corrected the redshifts of galaxies for the motion relative to the CMB and 
computed the co-moving distances \citep{2002sgd..book.....M} of galaxies.

Galaxy groups were  determined using the Friends-of-Friends cluster analysis 
method introduced in cosmology by \citet{1982Natur.300..407Z} and 
\citet{1982ApJ...257..423H}.
A galaxy belongs to a 
group of galaxies if this galaxy has at least one group member galaxy closer 
than a linking length. In a flux-limited sample the density of galaxies slowly 
decreases with distance. To take this selection effect into account properly 
when constructing a group catalogue from a flux-limited sample, we rescaled the 
linking length  with distance, calibrating the scaling relation by observed 
groups \citep[see ][for details]{2008A&A...479..927T, 2010A&A...514A.102T}. 
As a result, the 
maximum sizes in the sky projection and the velocity dispersions of our groups 
are similar at all distances. Our initial sample is chosen in the 
distance interval 120~\Mpc\ $\le d \le $ 340~\Mpc\ (the redshift range $0.04 < z 
< 0.09$) where the selection effects are the smallest \citep[we discuss the 
selection effects in detail in ][]{2010A&A...514A.102T, 2012A&A...542A..36E}.
The details of data 
reduction and the description of the group catalogue can be found in 
\citet{2012A&A...540A.106T}.

We calculate the galaxy luminosity density field to reconstruct the underlying 
luminosity distribution and to determine superclusters 
(extended systems of galaxies) in the luminosity density field
at smoothing length 8~\Mpc\ using $B_3$ spline kernel
\citep[see][for comparison between $B_3$ and
Gaussian kernel]{2007A&A...476..697E}. 
We created a set 
of density contours by choosing a density thresholds and defined connected 
volumes above a certain density threshold as superclusters. In order to choose 
a proper density level to determine individual superclusters, we analysed the 
properties of the density field superclusters at a series of density levels. 
As a result we used 
the density level $D_8 = 5.0$
(in units of mean density, $\ell_{\mathrm{mean}}$ = 
1.65$\cdot10^{-2}$ $\frac{10^{10} h^{-2} L_\odot}{(\vmh)^3})$
to determine individual superclusters
\citep{2012A&A...539A..80L}.
At this density level superclusters in the richest 
chains of superclusters in the volume under study  still form separate systems; 
at lower density levels they join into huge percolating systems. At higher 
threshold density levels superclusters are smaller and their number decreases. 

The  morphology of all superclusters used in this study was determined  
in \citet{2012A&A...542A..36E} applying Minkowski functionals.  
For a given surface the four Minkowski 
functionals are proportional to the enclosed 
volume $V$, the area of the surface $S$, the integrated mean curvature $C$, and 
the integrated Gaussian curvature $\chi$ 
\citep[see Appendix~\ref{sect:MF} and ]
[for details and references]{2007A&A...476..697E, 2011A&A...532A...5E}.
The overall morphology of a supercluster is described by the  shapefinders $K_1$ 
(planarity) and $K_2$ (filamentarity), and their ratio, $K_1$/$K_2$ (the shape 
parameter), calculated using the first three Minkowski functionals.  
The lower the value of the shape parameter, the more elongated a supercluster is.
The maximum value of the fourth Minkowski functional $V_3$ 
characterises the inner structure of the superclusters. The higher the 
value of $V_3$, the more complicated the inner morphology of a supercluster is
\citep{2007A&A...476..697E, 2011ApJ...736...51E}. Superclusters were
classified as filaments and spiders on the basis of their morphological
information and visual appearance.
Superclusters show wide morphological variety in 
which \citet{2011A&A...532A...5E} determined four main morphological types: 
spiders, multispiders, filaments, and multibranching filaments. Spiders and 
multispiders are systems of one or several high-density clumps with a number of 
outgoing filaments connecting them.  
Filaments and multibranching filaments are superclusters 
with filament-like main bodies that connect clusters in superclusters. 
For simplicity, in this study we 
classify superclusters as spiders and filaments. 
In Appendix~\ref{sect:MF} we show an examples of filament and spider-type
superclusters.

The description of the supercluster catalogues is given 
in \citet{2012A&A...539A..80L}, the details of supercluster
morphology are given in \citet{2007A&A...476..697E, 2011A&A...532A...5E}.

\begin{figure}[ht]
\centering
\resizebox{0.48\textwidth}{!}{\includegraphics[angle=0]{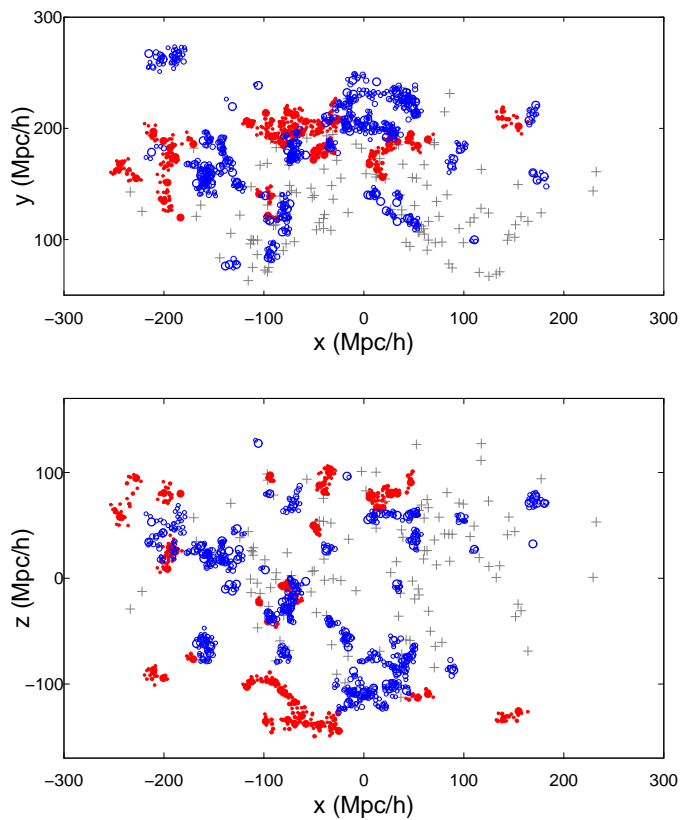}}
\caption{
Distribution  of galaxy groups with at least
four member galaxies in superclusters in $x$, $y$, and $z$ coordinates.
Red filled circles correspond to groups in filament-type superclusters.
Blue empty circles denote groups in spider-type superclusters.
Grey crosses denote groups with at least 30 member galaxies from the field.
}
\label{fig:sclxyz}
\end{figure}

Figure~\ref{fig:sclxyz} presents
the distribution of rich galaxy clusters in superclusters
of filament and spider morphology, as well as from the field in
cartesian coordinates $x$, $y$, and $z$ defined as in
\citet{2007ApJ...658..898P} and in \citet{2012A&A...539A..80L}:
\begin{equation}
\begin{array}{l}
    x = -d \sin\lambda, \nonumber\\[3pt]
    y = d \cos\lambda \cos \eta,\\[3pt]
    z = d \cos\lambda \sin \eta, 
\end{array}
\label{eq:xyz}
\end{equation}
where $d$ is the finger of god
corrected comoving distance, and $\lambda$ and $\eta$ are the SDSS 
survey coordinates.

\begin{figure}[ht]
\centering
\resizebox{0.48\textwidth}{!}{\includegraphics[angle=0]{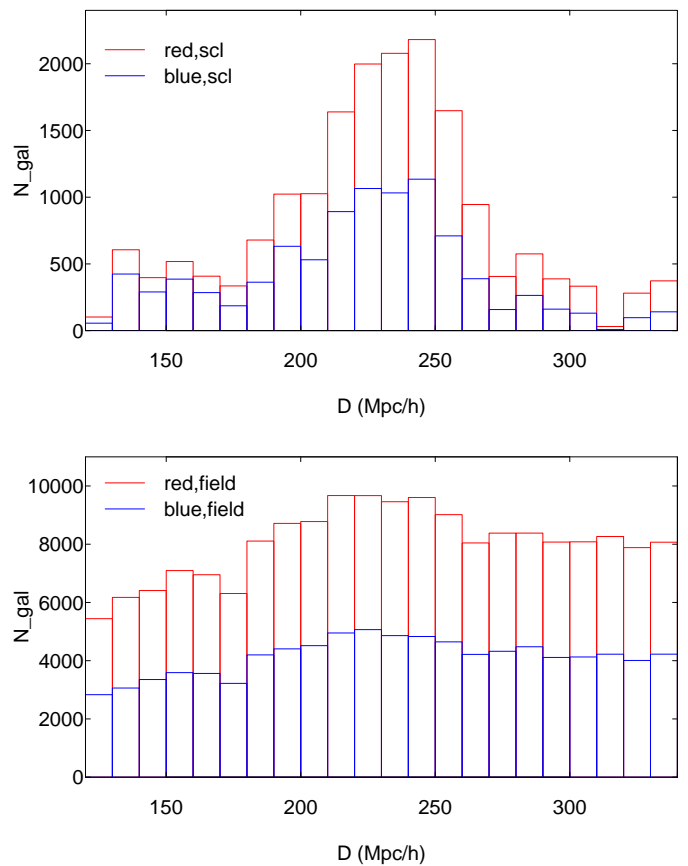}}
\caption{
The distribution  of distances of red and blue galaxies 
(red and blue lines, respectively) in superclusters (upper panel)
and in the field (lower panel).
}
\label{fig:redbludist}
\end{figure}

Up to distances of about 180~\Mpc\ the SDSS survey crosses the void region
between the Hercules supercluster and the chain of rich superclusters
farther away \citep[the Bootes void, ][]{1981ApJ...248L..57K}. 
Two chains of galaxies, groups, and superclusters
cross this void. At the distance interval of about 180~\Mpc\ $\le d \le $ 270~\Mpc\ 
the survey crosses a number of rich superclusters, including the Sloan Great
Wall \citep[detailed description of the large-scale distribution
of superclusters can be found in][]{2011ApJ...736...51E, 2011A&A...532A...5E}. 
At distances $d > $~270~\Mpc\ there is another void region, and
the frequency of superclusters in this region is very low. 
This is also seen in Fig.~\ref{fig:redbludist} where we plot
the distributions of distances to red and blue galaxies 
divided by their colour index $g - r = 0.7$ (see Sect.~\ref{subsec:galprop})
in superclusters and in the field.

We have different
distance intervals with different large-scale distributions and richness of systems.
From these we analysed data from two regions: 
a void region with poor systems, 120~\Mpc\ $\le d \le $ 180~\Mpc\ (near sample),
and a supercluster region with rich systems of galaxies 
180~\Mpc\ $\le d \le $ 270~\Mpc\ (distant sample). 
Among the 50 superclusters studied, 35 are of spider
morphology and 15 of filament morphology. 
The supercluster data are given in Table~\ref{tab:scldata}. 
We do not use data for very poor superclusters without any rich clusters
whose morphology is difficult
to determine \citep[for details we refer to][]{2012A&A...542A..36E}.

Our data are based
on a flux-limited sample of galaxies
in which absolute magnitude limit increases with distance,
introducing selection effects: there are relatively more red galaxies 
at great distances (this can also be seen in
Fig.~\ref{fig:redbludist}).  
In order to avoid this selection effect  
we used volume-limited samples
with $M_r \leq -18.5$  in near samples, and
$M_r \leq -19.5$ in distant samples.  
The numbers of galaxies 
in near and distant samples of superclusters are given in 
Table~\ref{tab:galdata}. The number of galaxies in filament-type superclusters
from near sample is small. Our calculations showed that
the results for the near and distant samples are similar, 
but the statistics of the near sample are not very reliable, 
therefore in the paper we present our results for
the distant sample only.

\begin{table}[ht]
\caption{Number of galaxies in filament and spider-type 
superclusters and in field for flux- and volume-limited samples.}
\begin{tabular}{rrrrr} 
\hline\hline  
           & \multicolumn{2}{c}{$120 - 180$~\Mpc}& \multicolumn{2}{c}{$180 - 270$~\Mpc}\\
 Sample  & flux-lim   & vol-lim & flux-lim  & vol-lim \\
\hline
 Filaments &   517    &   500  &    8102    &  6664 \\
 Spiders   &  3497    &  3024  &   12030    &  9211 \\
 Field     & 58712    & 50643  &  123230    & 93795 \\
\hline
                                                                             
\label{tab:galdata}                                                          
\end{tabular}\\
\end{table}

\subsection{Galaxy properties}
\label{subsec:galprop}

In the present paper we use the $g - r$ colours of galaxies and their absolute 
magnitudes in the $r$-band $M_r$. The absolute magnitudes of individual galaxies 
are computed according to the formula
\begin{equation}
M_r = m_r - 25 -5\log_{10}(d_L)-K,
\end{equation} 
 where $d_L$ is the luminosity distance in units of $h^{-1}$Mpc and
 $K$ is the $k$+$e$-correction. 
 The $k$-corrections were calculated with the \mbox{KCORRECT\,(v4\_2)} algorithm 
 \citep{2007AJ....133..734B} and the evolution corrections have been calibrated 
 according to \citet{2003ApJ...592..819B}. 
We use $M_{\odot} = 4.53$ (in $r$-filter). 
All of our 
magnitudes and colours correspond to  the rest-frame at  redshift $z=0$. 
The value $g - r = 0.7$ is used to separate red and blue galaxies, red galaxies 
having $g - r \geq 0.7$.
We used the morphology classification  by \citet{2011A&A...525A.157H}
which gives for each galaxy a probability of being early 
and late type, $p_e$ and $p_l$, correspondingly.
Approximately, we call a galaxy late type, if $p_l > 0.5$,
and  early type, if  $p_e > 0.5$.

To study the stellar masses $\log M_\mathrm{s}$ and SFRs, we
use the MPA-JHU spectroscopic catalogue \citep{2004ApJ...613..898T, 
2004MNRAS.351.1151B}. In this catalogue the different properties of 
galaxies are obtained by fitting SDSS photometry and spectra with
the stellar population synthesis models developed by \citet{2003MNRAS.344.1000B}.
The stellar masses of galaxies are derived in the
manner described by \citet{2003MNRAS.341...33K}. For the DR8 data
the stellar masses are estimated from the 
galaxy photometry \citep[rather than the spectral indices 
$D_n$4000 and $H_{\delta}$ 
used by][]{2003MNRAS.341...33K}. The SFRs 
are computed using the photometry and emission lines as described 
by \citet{2004MNRAS.351.1151B} and \citet{2007ApJS..173..267S}. 
For active galactic nuclei and 
galaxies with weak emission lines, SFRs are estimated from the 
photometry. For star-forming galaxies with strong emission lines, 
the SFRs are estimated by fitting different emission lines in the 
galaxy spectrum ($H_{\alpha}$, $H_{\beta}$, [\ion{O}{iii}] 5007, 
[\ion{N}{ii}] 6584, [\ion{O}{ii}] 3727, [\ion{S}{ii}] 6716).
The stellar masses and SFRs are taken from the SDSS CAS database.

In the group catalogue the main galaxy of a group is defined
as the most luminous galaxy in the $r$-band. We also use this definition 
in the present paper. In virialized clusters galaxies 
follow the cluster potential well, and  so we 
would expect that the main galaxies in clusters lie at the centres of 
groups and have low peculiar velocities 
\citep{1975ApJ...202L.113O}. Therefore the peculiar velocity  of the 
main galaxies in clusters is also an indication of the 
dynamical state of the cluster \citep{2009AJ....137.4795C, 2012A&A...540A.123E}. 
We calculate the peculiar velocities of the main galaxies, 
$|v_{\mathrm{pec}}|$, and compare these velocities for groups
in superclusters of different morphology, and in the field.

\section{Results}
\label{sect:results} 

\subsection{Superclusters of different morphology}
\label{sect:filspi} 

We analyse galaxy populations in superclusters of different morphology, 
and in the field using probability density  distributions 
which are 
calculated within the statistical package R environment 
using the 'density' command in the 'stats' 
package \citep{ig96} 
\footnote{\texttt~{http://www.r-project.org}}. 
In Tables in Appendix~\ref{sect:quantiles} we present the values of 
quantiles of the parameters analysed in the paper. 
The distributions of colours, probabilities of being 
early or late type, stellar masses, 
and SFRs are bimodal or asymmetrical; the use of
full distributions in the analysis and 
quantile values in Tables is straightforward. 
To not to make tables too
overcrowded we do not include errors to the tables;
the statistical significance of the differences 
between different galaxy properties have been found
using the full data (the integral distributions) and 
the Kolmogorov-Smirnov (KS) test. 
We give the $p_\mathrm{KS}$-values of the test, throughout 
the paper \citep[for details of this approach we refer to ][]{2008ApJ...685...83E}. 
We consider that the differences between distributions are highly significant
if the $p_\mathrm{KS}$ value (the estimated probability of rejecting the hypothesis
that distributions are statistically similar) $p_\mathrm{KS} \leq 0.05$.

\begin{figure}[ht]
\centering
\resizebox{0.48\textwidth}{!}{\includegraphics[angle=0]{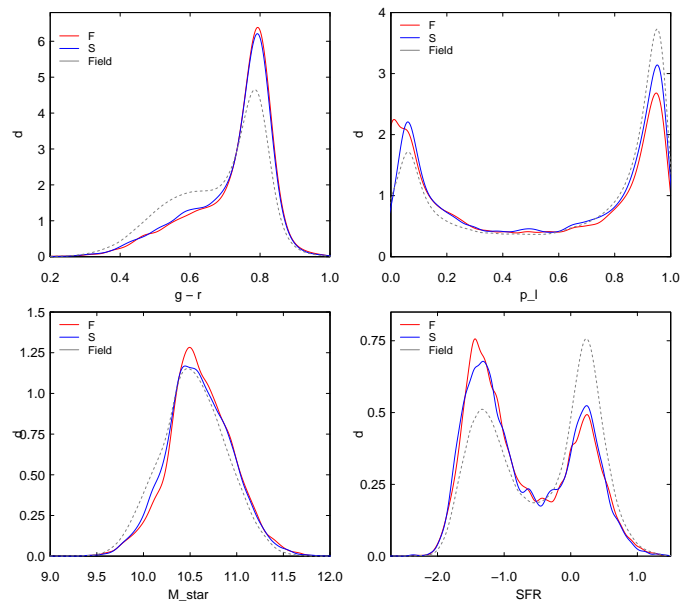}}
\caption{
Probability density distributions of galaxy $g - r$ colours (upper
left panel), probabilities of being late type ($p_l$, upper right panel), 
stellar masses ($\log M_\mathrm{s}$ in log $M_\odot$, 
lower left panel), and star formations rates (SFRs, in log $M_\odot$/yr) 
for galaxies in the superclusters of filament and spider morphology
and in the field.}
\label{fig:sclisodist}
\end{figure}

Figure~\ref{fig:sclisodist} present
the distributions of galaxy $g - r$ colours, 
probabilities of being late type ($p_l$), 
stellar masses ($\log M_\mathrm{s}$), and SFRs for galaxies from 
superclusters of filament and spider morphology, and from the field. 
We show the distributions of late-type galaxies only;
the distributions of early-type galaxies show similar
differences between galaxy populations ($p_e = 1 - p_l$).
The values of quantiles of galaxy parameters are given in 
Table~\ref{tab:filspiisopara}.

This figure shows, first of all, that galaxy populations in superclusters
and in the field are  different, superclusters containing a larger fraction
of red, early-type, high stellar mass, and low SFR galaxies
than can be found in the field. These figures show that
galaxy populations in superclusters of filament and spider morphology are also 
different, superclusters of filament morphology contain a larger fraction
of red, early-type, low SFR galaxies than superclusters of spider
morphology. All differences are statistically significant
at very high confidence level
($p_\mathrm{KS} < 0.001$).

\begin{figure*}[ht]
\centering
\resizebox{0.90\textwidth}{!}{\includegraphics[angle=0]{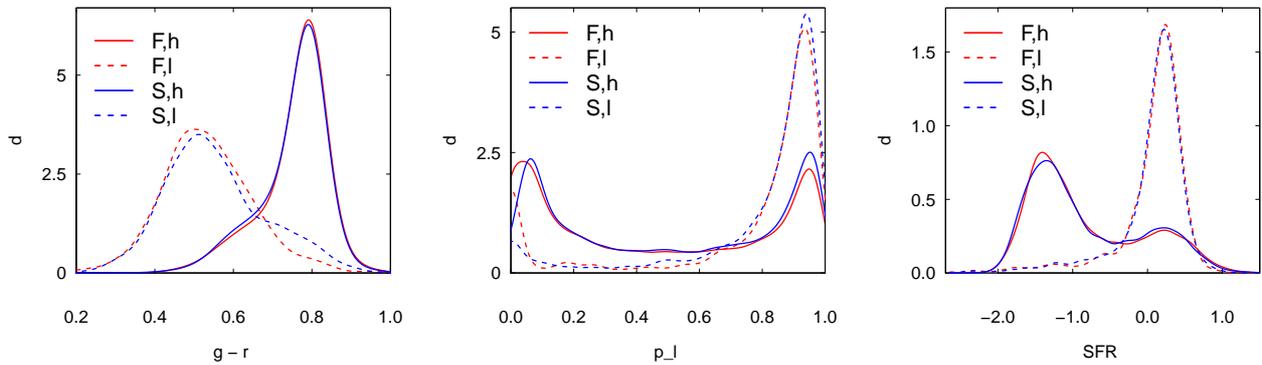}}
\caption{
Probability density distributions of galaxy $g - r$ colours (left
panel), probabilities of being late type ($p_l$, middle panel), 
and star formations rates (SFRs, in log $M_\odot$/yr, right panel) 
for high (h, $log M_s \geq 10.25$, solid lines) and 
low (l, $log M_s < 10.25$, dashed lines) stellar mass galaxies 
in the superclusters of filament (F, red lines) and spider 
(S, blue lines) morphology.}
\label{fig:filspihilo}
\end{figure*}

The distribution of stellar masses of galaxies in filament and spider-type 
superclusters shows that in filament-type superclusters there is some deficit of 
low stellar mass galaxies with $\log M_s < 10.25$, and relatively more
galaxies of higher stellar mass than in superclusters of spider morphology.
In Fig.~\ref{fig:filspihilo} we show the probability density distributions 
of colours, types, and SFRs of high and low stellar mass galaxies
in superclusters of filament and spider type. 
Table~\ref{tab:filspiisohilosmass}
presents the values of quantiles of these parameters. 
As expected \citep[see, for example, ][]{2009ARA&A..47..159B}, 
among high stellar mass galaxies there 
are relatively more
red, early-type, low SFR galaxies than among low stellar
mass galaxies. 
Interestingly, this figure
shows that low stellar mass galaxies in filament-type superclusters are mainly 
blue, but in spider-type superclusters there is relatively more red galaxies 
among them in comparison with filament-type superclusters.   
These differences are statistically highly significant, as also differences in 
galaxy types. 
Differences between SFRs of galaxies from filament and spider-type
superclusters divided by stellar mass are low and statistically 
unsignificant, according to the KS test. 

\begin{figure}[ht]
\centering
\resizebox{0.48\textwidth}{!}{\includegraphics[angle=0]{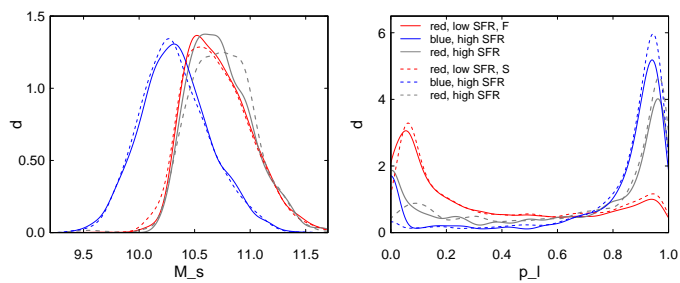}}
\caption{
Probability density distributions of 
stellar masses ($\log M_\mathrm{s}$, in $\log M_\odot$), 
(left panel) and probabilities of being late type ($p_l$, right panel), 
for red, low SFR galaxies (red lines), blue, high SFR galaxies (blue lines),
and for red, high SFR galaxies (grey lines)
in the superclusters of filament (solid lines) and spider (dashed lines) morphology.
High SFR galaxies have $\log \mathrm{SFR} \geq -0.5$, and low SFR galaxies have $\log SFR < -0.5$.
Lines in the left panel denote the same populations as in the right panel. }
\label{fig:filspipassact}
\end{figure}

Next we divided galaxies by colours and SFRs as red, low SFR galaxies
($g - r \geq 0.7$, and $\log \mathrm{SFR} < -0.5$, approximately 60\% of all galaxies), blue, high SFR galaxies
($g - r < 0.7$, and $\log \mathrm{SFR} \geq -0.5$, 30\% of all galaxies), and red, high SFR galaxies
($g - r \geq 0.7$, and $\log \mathrm{SFR} \geq -0.5$, 10\% of all galaxies). 
We plot the distributions
of their stellar masses and types in Fig.~\ref{fig:filspipassact},
and give the values of quantiles of the stellar masses and types
in Table~\ref{tab:filspihilosfr}.
This figure shows interesting feature: stellar masses of red, low SFR galaxies,
and blue, high SFR galaxies in filament-type superclusters are higher than
in spider-type superclusters, as were these masses in overall distributions
(Fig.~\ref{fig:sclisodist}), but stellar masses of red, high SFR galaxies
have higher values in spider-type superclusters. 
In filament-type superclusters their stellar masses are close to
the stellar masses of red, low SFR galaxies, in spider-type superclusters
their stellar masses are even higher than those of red, low SFR galaxies. 
These galaxies are mostly of late type. Thus this figure shows that 
high mass galaxies are red, even if they are of high SFR 
and of late type, and there is a difference 
in their masses in different types of superclusters.  
KS test shows that the differences between
stellar masses and types of high and low SFR 
galaxies from filament and spider-type superclusters 
are statistically highly significant for both high and low SFR red galaxies.

\begin{figure}[ht]
\centering
\resizebox{0.48\textwidth}{!}{\includegraphics[angle=0]{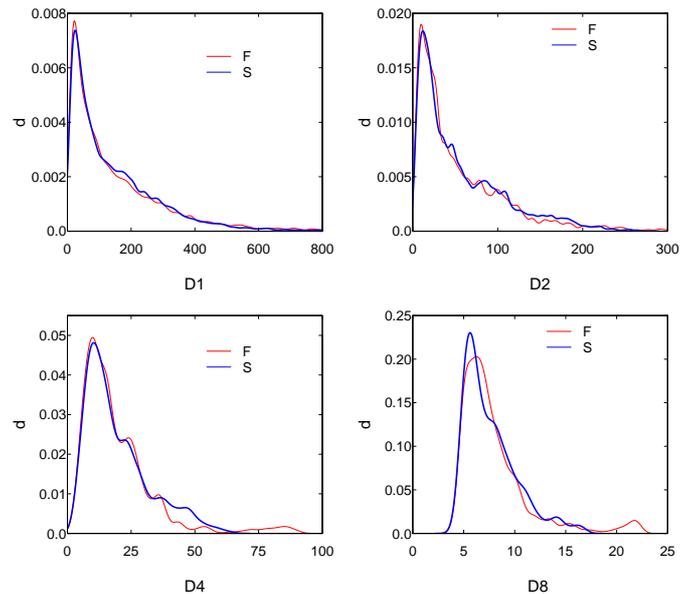}}
\caption{
Distributions of environmental densities 
at different smoothing lengths (1, 2, 4, and 8~\Mpc) 
 in superclusters of filament (red lines)
and spider (blue lines) morphology. 
}
\label{fig:filspiden}
\end{figure} 

Figure~\ref{fig:filspiden} shows the distributions of environmental densities
at smoothing lengths 1, 2, 4, and 8~\Mpc\ in filament and spider-type superclusters.
At all smoothing lengths in these distributions poor groups with $N_\mathrm{gal} \leq 9$
dominate at low densities; higher
values of densities correspond to richer groups. 
Figure~\ref{fig:filspiden} shows a small
deficit of intermediate-value densities in filament-type superclusters, the KS test
shows that the differences are statistically significant at least 95\% level.
If densities in filament-type superclusters were systematically higher than in
spider-type superclusters then the differences between galaxy populations 
in these superclusters 
were related  to the dependence of the galaxy morphology on the environmental density, 
but
we see that this is not so. Inner morphology of superclusters is important.

\begin{figure}[ht]
\centering
\resizebox{0.45\textwidth}{!}{\includegraphics[angle=0]{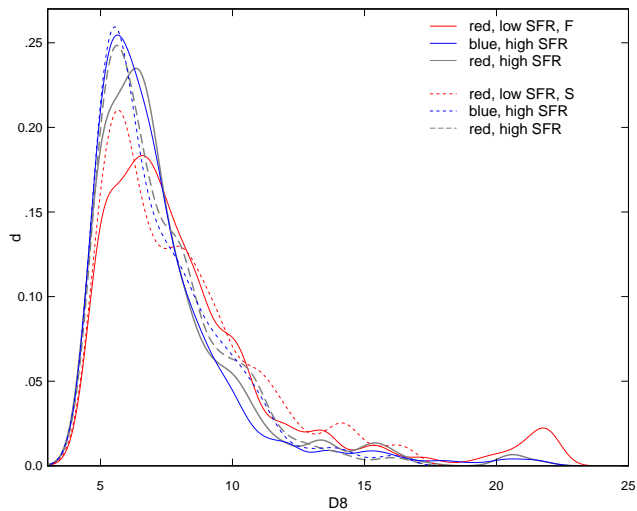}}
\caption{
Distributions of environmental density 
at smoothing length 8~\Mpc\ around red, low SFR galaxies
(red lines), blue, high SFR galaxies (blue lines), and red, high SFR galaxies
(grey lines) in superclusters of filament (solid lines)
and spider (dashed lines) morphology. 
}
\label{fig:filspirgbden}
\end{figure} 

We also calculated environmental densities around galaxies of different colour
and SFRs. Figure~\ref{fig:filspirgbden} shows environmental densities at 8~\Mpc\ smoothing 
length around red, low SFR, blue, high SFR, and red, high SFR galaxies, analysed above.
In this figure we see that blue, high SFR galaxies have somewhat lower environmental 
densities around them than red, low SFR galaxies in both type of superclusters.
This is an evidence of a large-scale morphology-density relation.  
Environmental densities around red, low SFR galaxies in filament-type superclusters
have higher values than in spider-type superclusters. Environmental densities
around red, high SFR galaxies in spider-type superclusters are close to those 
for blue galaxies; in filament-type superclusters the densities around them
have somewhat higher values. In filament-type superclusters some of these galaxies
are located in a rather high-density environment. All differences are statistically highly
significant. 

Our sample contain several very rich superclusters. We analyse their galaxy populations
separately (Sect.~\ref{sect:4scl}). To see whether our overall results are affected 
by their  dominance we repeated all calculations excluding the richest
systems from the sample. All our results remained the same, therefore
neither the superclusters from the Sloan Great Wall, nor other very rich
superclusters do not dominate in our results.

\subsection{Superclusters of different shape}
\label{sect:k12}

Superclusters can be divided into more and less elongated systems
according to the value of their shape parameter $K_1$/$K_2$.  
The smaller the shape parameter, the more elongated a supercluster is.
Superclusters with  $K_1/K_2 < 0.5$ (more elongated systems) are 
called  filament type,
and those with $K_1/K_2 \geq 0.5$ belong to pancake type being less elongated
systems \citep{2011A&A...532A...5E}. 
\citet{2013MNRAS.428..906C} analysed stellar populations of galaxies
in superclusters of different shape, using shape parameter to characterise
the overall morphology of superclusters. Similarly, we use
the shape parameter to divide superclusters
of filament and spider types into sets of more elongated and less
elongated systems and compare galaxy populations in them. The results 
are presented with the KS test in Table~\ref{tab:ks4scl}.  
For volume-limited samples galaxy populations
in more elongated and less elongated
superclusters of filament morphology are statistically similar.
KS test shows that galaxy types and SFRs in more
elongated and less elongated superclusters of spider type are 
different at very high significance level. \citet{2013MNRAS.428..906C} 
did not find differences
in galaxy populations in superclusters of different shape, and we found this
for spider-type systems only. We analysed quite rich systems only, diving them
at first by their inner morphology. This may lead to the differences in our 
results.

\subsection{The richest superclusters}
\label{sect:4scl} 
 
\begin{figure}[ht]
\centering
\resizebox{0.48\textwidth}{!}{\includegraphics[angle=0]{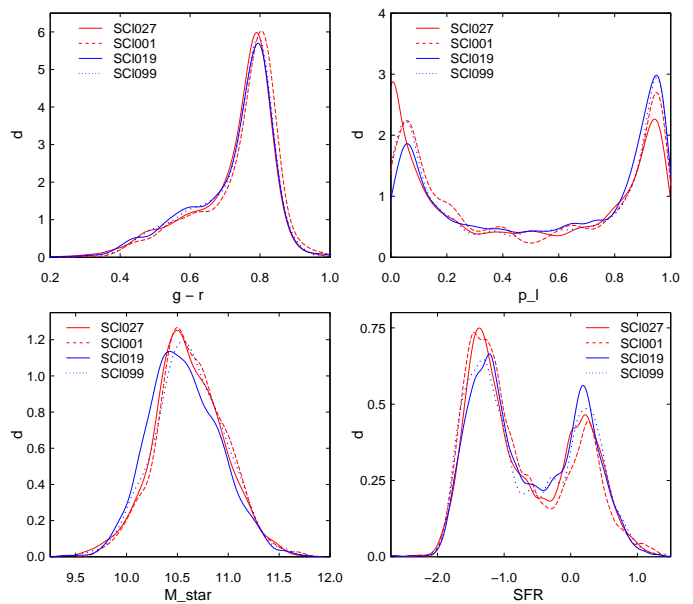}}
\caption{
The same as in Fig.~\ref{fig:sclisodist} for the richest superclusters:
filaments SCl~001 and SCl~027 (red dashed and solid lines), and 
spiders SCl~019 and SCl~099 (blue solid and dashed lines).
}
\label{fig:4scl}
\end{figure}

Next we compared galaxy populations in the richest individual superclusters
of both morphological types, the superclusters of filament morphology, 
SCl~001 and SCl~027, and the superclusters of spider morphology, SCL~019 and
SCl~099. The superclusters 
SCl~027 and SCl~019 are the richest systems in the Sloan Great Wall
\citep{2011ApJ...736...51E}. The superclusters
SCl~099 (the Corona Borealis supercluster) and SCl~001 belong to the 
dominant supercluster plane \citep{1997A&AS..123..119E, 2011A&A...532A...5E}.
Figure~\ref{fig:4scl} shows galaxy populations in these superclusters,
and Table~\ref{tab:ks4scl} the results of the KS test, where we compare
galaxy content of filament-type systems and 
spider-type systems. 

The galaxy populations in the individual richest
systems are all different. Comparison of the galaxy populations
of the richest filament-type superclusters 
and spider-type superclusters shows that the distributions
of all their galaxy properties 
considered in this paper are different at very high significance level,
the fractions of red, low SFR 
galaxies in the richest filament-type systems are higher than those in
the richest spider-type systems. For the richest filament-type systems
Table~\ref{tab:ks4scl} shows that all their galaxy properties except
stellar masses are different at very high significance level, the fraction
of red galaxies in the SCl~001 is higher than in SCl~027, and the fraction
of galaxies with high SFR is lower. For the richest spider-type superclusters
only the stellar masses of galaxies in them are different at statistically
high significance level, galaxies in SCl~099 having higher stellar masses
than in those SCl~019.

\begin{table}[ht]
\caption{The results of the KS test (the $p_\mathrm{KS}$-values): comparison
of the galaxy properties in filament- and spider-type 
superclusters of different overall shape and richness.}
\begin{tabular}{rrrrr} 
\hline\hline  
(1)&(2)&(3)&(4)&(5)\\      
\hline 
             \multicolumn{5}{c}{More elongated and less elongated superclusters}\\
\hline 
            & \multicolumn{2}{c|}{Filaments}& \multicolumn{2}{c}{Spiders}\\
\hline
         &  \multicolumn{1}{c|}{flux-lim}   & \multicolumn{1}{c|}{vol-lim} & \multicolumn{1}{c|}{flux-lim}  & \multicolumn{1}{c}{vol-lim} \\
         &     \multicolumn{1}{c|}{$p_\mathrm{KS}$     }&     \multicolumn{1}{c|}{$p_\mathrm{KS}$  } &   \multicolumn{1}{c|}{$p_\mathrm{KS}$}   &  $p_\mathrm{KS}$   \\
\hline
$g - r$  &  \multicolumn{1}{c|}{0.515     }&  \multicolumn{1}{c|}{0.735} &  \multicolumn{1}{c|}{0.331} &   0.107  \\
$p_l$    &  \multicolumn{1}{c|}{0.099     }&  \multicolumn{1}{c|}{0.535} &  \multicolumn{1}{c|}{0.003} &   0.011  \\
$\log M_\mathrm{s}$    &  \multicolumn{1}{c|}{$<1e-5$}&  \multicolumn{1}{c|}{0.153} &  \multicolumn{1}{c|}{0.524} &   0.350 \\
$SFR$    &  \multicolumn{1}{c|}{0.242     }&  \multicolumn{1}{c|}{0.221} &  \multicolumn{1}{c|}{0.001} &   0.002  \\
                                                                                               
\hline 
\\
              \multicolumn{5}{c}{The richest superclusters}\\
\hline 
              & \multicolumn{2}{c|}{Filaments}& \multicolumn{2}{c}{Spiders}\\
              & \multicolumn{2}{c|}{SCl~001, SCl~027}& \multicolumn{2}{c}{SCl~019, SCl~099}\\
\hline
         &  \multicolumn{1}{c|}{flux-lim}   & \multicolumn{1}{c|}{vol-lim} & \multicolumn{1}{c|}{flux-lim}  & \multicolumn{1}{c}{vol-lim} \\
         &  \multicolumn{1}{c|}{$p_\mathrm{KS}$     } &  \multicolumn{1}{c|}{$p_\mathrm{KS}$     } &  \multicolumn{1}{c|}{$p_\mathrm{KS}$     }  &    $p_\mathrm{KS}$        \\
\hline
$g - r$  &  \multicolumn{1}{c|}{$<1e-5$}&   \multicolumn{1}{c|}{$<1e-5$} & \multicolumn{1}{c|}{0.824     }&  0.804    \\
$p_l$    &  \multicolumn{1}{c|}{$<1e-5$}&   \multicolumn{1}{c|}{$<1e-5$} & \multicolumn{1}{c|}{0.001     }&  0.127    \\
$\log M_\mathrm{s}$    &  \multicolumn{1}{c|}{$<1e-5$}&   \multicolumn{1}{c|}{0.063     } & \multicolumn{1}{c|}{0.012     }&  $<1e-5$\\
$SFR$    &  \multicolumn{1}{c|}{0.001     }&   \multicolumn{1}{c|}{0.003     } & \multicolumn{1}{c|}{0.003     }&  0.079    \\
$D_8$    &  \multicolumn{1}{c|}{$<1e-5$}&   \multicolumn{1}{c|}{$<1e-5$} & \multicolumn{1}{c|}{$<1e-5$}&  $<1e-5$ \\
\hline
                                                                                                   
\label{tab:ks4scl}                                                          
\end{tabular}\\
\tablefoot{                                                                                 
Columns are as follows:
1: Galaxy properties. $g - r$ -  colour index, 
$p_l$ - probability of being a late type, 
$\log M_\mathrm{s}$ - log stellar mass, $SFR$ - log SFR, $D_8$ - environmental density
at 8 \Mpc\ scale smoothing scale;
2--5: the $p_\mathrm{KS}$-value of the test. 
}
\end{table}

\subsection{Rich and poor groups in superclusters}
\label{sect:nggr} 

\begin{figure}[ht]
\centering
\resizebox{0.48\textwidth}{!}{\includegraphics[angle=0]{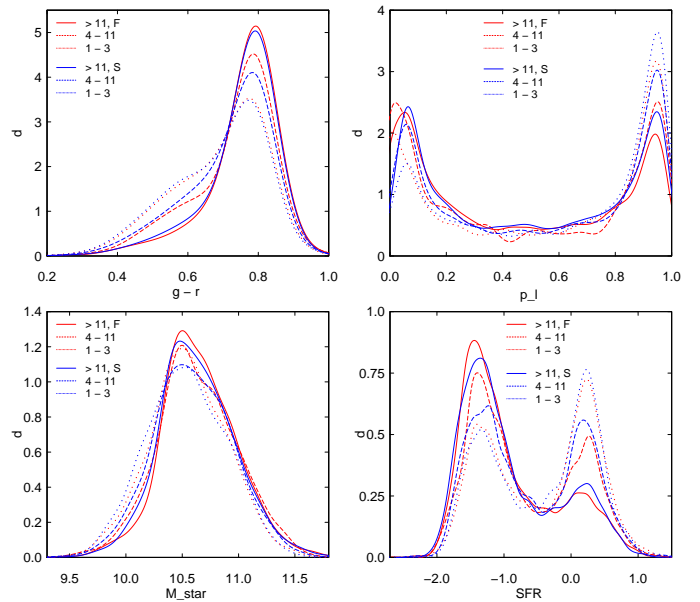}}
\caption{
The same as in Fig.~\ref{fig:sclisodist} for groups of different richness
in superclusters of filament and spider morphology (red and blue lines). 
Solid lines show distributions for groups with at least 12 member galaxies.
Dashed lines correspond to distributions 
for groups with $4 \leq N_\mathrm{gal} \leq 11$.
Dotted lines to single galaxies and groups with $N_\mathrm{gal} \leq 3$.
}
\label{fig:gr14}
\end{figure}

Most galaxies are located in groups of various richness. Therefore,
as a next step we compared the galaxy content of groups of different richness
in filament- and spider-type superclusters and in the field. We divided groups
of galaxies in superclusters and in the field according to their richness
as very poor systems with $N_\mathrm{gal} \leq 3$, poor groups with the number
of member galaxies $4 \leq N_\mathrm{gal} \leq 11$, and rich groups with 
$N_\mathrm{gal} \geq 12$. The comparison of galaxy properties in three richness
classes are given in Fig.~\ref{fig:gr14}. For clarity
we present in Fig.~\ref{fig:gr14} 
the results for groups from superclusters only.
Table~\ref{tab:filspinggr}
presents the values of quantiles of galaxy properties. 

Figure~\ref{fig:gr14} shows that  the richer the group,
the larger is the fraction of red, early-type, low SFR,
high stellar mass galaxies in it. We found a similar trend  for
groups in
low-density global environments. The KS
test shows that differences between all galaxy parameters
from poorer and richer groups are statistically highly significant. 

Figure~\ref{fig:gr14} shows that
colour distributions of galaxies in very poor groups  
with $N_\mathrm{gal} \leq 3$
are very similar in superclusters of different type. 
The KS test {shows that the differences
are not statistically significant}
(Table~\ref{tab:ksn430}). Galaxies in very poor 
groups in superclusters of filament-type have higher stellar masses than galaxies
in superclusters of spider-type, and the fraction of high SFR galaxies
among them is smaller.
For richer groups Fig.~\ref{fig:gr14} shows that 
groups of the same richness host galaxies with higher stellar masses, 
and they have a larger fraction of early-type and red galaxies, 
and a higher fraction of low SFR galaxies, if they are located 
in superclusters of filament morphology. For colours and SFRs the KS
test says that the differences between groups with less and 
more than 12 galaxies are not statistically significant. The stellar masses
and types of galaxies 
are also significantly different in the case of these richness classes.
In the field groups
the fraction of red, low SFR galaxies is lower than 
in groups of the same richness in superclusters.
Summarising, the galaxy content of groups of the same richness is somewhat different
in superclusters of filament and spider morphology.

\begin{table}[ht]
\caption{The results of the KS test (the $p_\mathrm{KS}$-values): galaxy properties in 
groups of different richness
in superclusters of filament and spider morphology.}
\begin{tabular}{rrrr} 
\hline\hline  
(1)&(2)&(3)&(4) \\      
\hline 
Property & $N_\mathrm{gal} < 4$ & $4 \leq N_\mathrm{gal} \leq 11$ & $N_\mathrm{gal} \geq 12$  \\
         & $p_\mathrm{KS}$       &  $p_\mathrm{KS}$&    $p_\mathrm{KS}$       \\
\hline
$g - r$  & 0.359     &  $9e-05$    &  0.122        \\
$p_l$    & $<1e-5$   &  $<1e-5$ &  $<1e-5$   \\
$\log M_\mathrm{s}$    & 0.006     &  0.017      &  0.014        \\
$SFR$    & 3e-04     &  0.022      &  0.388          \\
\hline
                                                               
\label{tab:ksn430}                                                          
\end{tabular}\\
\tablefoot{                                                                                 
Columns are as follows:
1: Galaxy properties are the same as in Table~\ref{tab:ks4scl}. 
2: the $p_\mathrm{KS}$-values of the test
for single galaxies and groups with 2-3 member galaxies in superclusters of filament and spider
morphology.
3: the same for groups with 4 - 11  member galaxies.
4: the same for groups with at least 12  member galaxies.

}
\end{table}

\subsection{The peculiar velocities of the main galaxies of groups}
\label{sect:pec} 

\begin{figure}[ht]
\centering
\resizebox{0.48\textwidth}{!}{\includegraphics[angle=0]{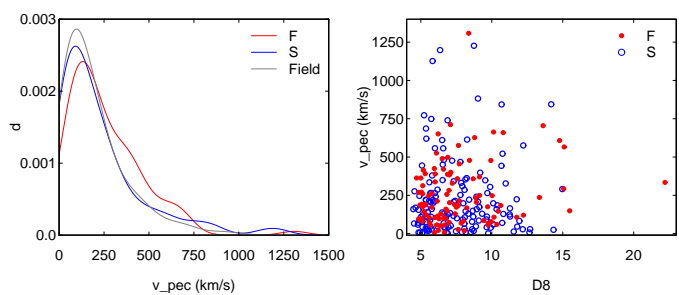}}
\caption{
Left panel: the peculiar velocities of the main galaxies in groups with
at least 12 member galaxies in superclusters of filament and spider morphology (red and blue lines), 
and in the field (grey line). 
Right panel: the peculiar velocities of the main galaxies in groups with
at least 12 member galaxies in superclusters of filament and spider morphology (red and blue dots) 
vs. environmental density in superclusters at smoothing length 8~\Mpc. 
}
\label{fig:pecvel}
\end{figure}

\citet{2012A&A...542A..36E} showed that in superclusters of spider morphology 
rich groups with $N_\mathrm{gal} \geq 50$ have 
a greater probability of having
substructure and higher peculiar velocities $|v_{\mathrm{pec}}|$ 
of their main galaxies
than rich groups in superclusters of filament morphology. They 
suggested that rich groups in superclusters of spider morphology
may be dynamically younger than rich groups in superclusters of filament 
morphology. Next we compare the peculiar velocities of the main galaxies
of groups in superclusters and in the field to obtain an estimate of
the group's dynamical state, and to understand whether the differences
in the dynamical state of groups may be related to the differences in
their galaxy content. 
In these calculations we did not want to use data of all groups since
for very poor groups calculations of the peculiar
velocities of their main galaxies are not reliable. 
Recently \citet{2013MNRAS.434..784R}
used several methods to analyse the structure and dynamical state of 
galaxy groups and showed that these methods 
give quite reliable results for groups with more than ten member galaxies 
\citep[Figure~2 and Table~1 in ][]{2013MNRAS.434..784R}. 
We made calculations with several group richness limits and found that
the results are similar, but for richer groups the sample size is smaller and 
this decreases the statistical significance of the results. Therefore  
we used groups with $N_\mathrm{gal} \geq 12$ in our analysis.
We show the distributions of the values of the peculiar velocities of the
main galaxies in groups in superclusters and in the field
in Fig.~\ref{fig:pecvel} (left panel).

Figure~\ref{fig:pecvel} shows that the distributions are clearly different.
The KS test says that the differences of the peculiar velocities
of the main galaxies in groups with at least 12 member galaxies
in superclusters of filament and spider morphology,
and between supercluster and field galaxy groups 
are statistically significant at very high levels (p-value $p < 0.05$).
There is a deficit of groups with small values of the peculiar velocities 
of their main galaxies in
groups from superclusters of filament morphology, and an excess
in groups from superclusters of spider morphology, and in the field.
\citet{2012A&A...540A.123E} showed for very rich groups with $N_\mathrm{gal} \geq 50$
that in groups with the peculiar velocities of the main galaxies 
$|v_{\mathrm{pec}}| < 250$ $km s^{-1}$
main galaxies are located near the centre of the group. 
Higher peculiar velocities of the 
main galaxies suggest that the main galaxy is located far from the centre, a 
sign of a dynamically young group. Figure~\ref{fig:pecvel} shows that
the peculiar velocities of the main galaxies in groups in superclusters
of filament morphology are higher than those in groups from superclusters of
spider morphology, suggesting that they are dynamically younger. 
In the richest groups we see the opposite: here  the peculiar velocities 
of the main galaxies in groups from spider-type superclusters are
higher than those in groups of filament-type superclusters,
as also shown in \citet{2012A&A...542A..36E}.
Interestingly, we found that groups of the same richness contain a larger fraction
of red, passive galaxies in superclusters of filament
morphology than in superclusters of spider morphology, and this trend is similar
for groups of different richness, but the values of the peculiar velocities
depend on richness. 

We also studied the relation between the  peculiar 
velocities of group main galaxies, and environmental density in superclusters
around them, for several smoothing lengths. Since the results were similar, we 
present them for smoothing length 8~\Mpc\ only (Fig.~\ref{fig:pecvel}, right panel).
This figure shows a possible weak correlation between the peculiar velocities
of the main galaxies and environmental density for filament-type superclusters,
but not for spider-type superclusters. This was confirmed by Pearson's 
correlation test which showed for filament-type superclusters
that the correlation coeficient $r = 0.19$ with $p = 0.06$.
For spider-type superclusters this test found that $r = 0.01$ with $p = 0.93$.

\section{Discussion and conclusions}
\label{sect:discussion} 

We showed that the fraction of red, early-type, 
low SFR galaxies in superclusters of filament morphology is higher 
than in superclusters of spider morphology. In addition, 
the fraction of these galaxies 
in superclusters is higher than among  galaxies in low-density global 
environments. 
There are relatively more red galaxies among low stellar mass galaxies 
in spider-type superclusters than in filament-type superclusters.

In superclusters of filament morphology
groups of equal richness host galaxies with higher stellar masses, 
and they have a larger fraction of early type and red galaxies, 
and a higher fraction of low SFR galaxies in comparison with groups
in spider-type superclusters.
In agreement with \citet{2012A&A...545A.104L}, we found
that groups of equal richness 
have a larger fraction of early-type and red galaxies, 
and a higher fraction of low SFR galaxies, if they are located 
in high-density large-scale environments (superclusters)
rather than in the field. In superclusters they also
host galaxies with higher stellar masses than groups in the field.
\citet{2013MNRAS.432.1367L} also found that galaxy groups in superstructures
\citep{2011MNRAS.415..964L}
have higher stellar masses, higher values of the peculiar velocities, and
have older galaxy populations than groups outside of superstructures.

About 10\% of all galaxies in both types of 
superclusters are red, high SFR galaxies. 
These galaxies are mostly late type,  their stellar masses are similar
to stellar masses of red, low SFR galaxies, and higher than those of blue
galaxies. In spider-type superclusters their stellar masses have higher values
than in filament-type superclusters. 
Earlier studies have shown that massive galaxies are red, independent of 
their morphological types 
\citep{2009MNRAS.393.1324B, 2010MNRAS.405..783M, 2011ApJ...736...51E}.
We found that red, high SFR galaxies can usually be found 
at low and intermediate
densities in both type of superclusters, but some of them are also located 
at regions of quite high environmental density.  
Red, high SFR galaxies typically lie at intermediate densities in small groups
in superclusters \citep{2005A&A...443..435W, 
2008ApJ...685...83E, 2009MNRAS.399..966S}, 
but they can be also found in rich clusters, and among the main 
galaxies of clusters \citep{2011ApJ...736...51E}, in agreement with our
present findings. This means that
processes which change the colours of late-type galaxies to red while 
keeping the high SFR must occur both in high- and intermediate-density
regions, and are more effective on high stellar mass galaxies.
We plan to study their location in groups and clusters of individual
superclusters to understand their distribution and properties better.

\citet{2008ApJ...685...83E, 2011ApJ...736...51E} determined a large
variation in the distributions of different galaxy populations
in individual rich superclusters. We found in the present study
a large diversity in the properties of 
galaxy populations of the richest superclusters in our sample.
Explaining large variation in the distribution of galaxy
populations in superclusters is a challenge for galaxy formation
models.

\citet{2013MNRAS.428..906C} showed that there is no difference between 
stellar populations of galaxies in superclusters classified by their overall shape
as filaments and pancakes. In our study 
we found that when we divided superclusters of different morphology
by their shape
parameter into more elongated and less elongated populations then
the differences in galaxy population between them are small. 

Superclusters obtained their  morphology and group and galaxy populations
during the formation and evolution of the cosmic web. 
In the $\Lambda$ Cold Dark Matter ($\Lambda$CDM) 
concordance cosmological model the structures forming the 
cosmic web grow by hierarchical clustering driven by gravity \citep[see ][and 
references therein]{2002PhRvD..65d7301L, 2008arXiv0804.2258L}. 
The present-day dynamical 
state of clusters of galaxies depends on their formation history. 
We analysed the dynamical state of galaxy groups using the peculiar
velocity of their main galaxies as an indicator of the dynamical state, 
and found that the velocities in filament-type superclusters are higher
than those in spider-type superclusters. In this analysis we used
groups with $N_\mathrm{gal} \geq 12$. In contrast, for the richest
clusters with $N_\mathrm{gal} \geq 50$ \citet{2012A&A...542A..36E} found that
these velocities are higher in superclusters of  spider morphology.

\citet{2013A&A...551A.143K} recently analysed the distribution and dynamical state
of galaxy groups in the Ursa Major Supercluster (in our catalogue SCl~211
of filament morphology) and found that 
groups with Gaussian velocity distribution of their member galaxies are located
in the denser regions of the supercluster. They suggested 
that relaxed galaxy groups in the supercluster have formed and evolved earlier 
and faster around high-density peaks, 
while nonrelaxed systems may be growing more slowly on the peripheries of lower 
density peaks. Interestingly, we found
in this study a weak correlation between environmental density
and the peculiar velocities of the main galaxies in groups in
filament-type superclusters, so that groups with higher values
of the peculiar velocities of their main galaxies are located in higher
density environments, this is opposite to the \citet{2013A&A...551A.143K}
result. They studied one rich supercluster.
We showed that
the properties of individual rich superclusters vary strongly. 
This may be the reason for different results.
\citet{2011ApJ...736...51E} found in the study of the richest superclusters
in the Sloan Great Wall, SCl~027 and SCl~019, that the peculiar velocities
of the main galaxies in groups depend on the morphology of the main galaxies,
and on their location in the supercluster. 

\citet{2012A&A...542A..36E} assumed that 
superclusters of spider morphology have richer inner structure than 
superclusters of filament morphology with a large number of filaments between 
clusters in them so mergers of clusters 
may occur more often in these superclusters. 
Therefore clusters in spider-type superclusters could be dynamically younger.
This agrees with the results of their galaxy populations,
spider-type superclusters hosting a larger fraction of blue, star forming
galaxies, but
contradicts the calculations of the peculiar velocities of their main
galaxies. This is one of the open questions for future studies of
superclusters and galaxy clusters in them.

\citet{2005A&A...439...45E} showed by numerical simulations that in 
low-density global environment only poor systems form, 
these systems grow very
slowly. In high-density global environments, the dynamical evolution is
rapid and starts earlier, consisting of a continua of transitions of dark
matter particles to building blocks of galaxies, groups, and
clusters. Dark matter halos in voids are found to be less
massive, less luminous, and their growth of mass is suppressed and
stops at earlier epochs than in high-density regions. 
During the structure formation in the Universe, 
halo sizes in the supercluster core regions increase,
while in void regions halo sizes remain 
unchanged \citep{2009A&A...495...37T}. Thus, during the evolution of structure 
the overall density in superclusters increases, 
which enhances the evolution of the small-scale protohalos in them
\citep{2011A&A...531A.149S}.
This may lead to the differences in the properties of galaxy
groups and galaxies in them in superclusters and in the field.

A number of processes that transform galaxies in high-density environments 
have been proposed, either observationally or in simulations. 
These include galaxy-galaxy mergers \citep{1992ARA&A..30..705B}, 
galaxy harassment \citep{1976ApJ...204..642R, 1996Natur.379..613M},
tidal stripping by the host group or cluster's 
gravitational potential \citep{2003ApJ...582..141G}, 
ram pressure stripping of cold gas by the hot gas of the
 diffuse intra-cluster or intra-group medium \citep{1972ApJ...176....1G}, 
and strangulation or removal of galactic halo gas
\citep{1980ApJ...237..692L}. These processes are effective both in galaxy groups
and during mergers of groups into clusters 
\citep[pre- and post-processing of galaxies,][]{2009ApJ...690.1292B,
2013MNRAS.435.2713V}.

\citet{2011ApJ...741...99C} used hydrodynamical simulations to show that 
for galaxies at $z=0$ star formation is efficient in low-density 
regions while it is substantially suppressed in cluster environments. 
During hierarchical structure formation, gas is heated in 
high-density regions (groups, clusters, and superclusters), 
and an increasingly larger fraction of the gas has entered 
a phase which is too hot to feed the residing galaxies. 
As a result, the cold gas supply to galaxies in these regions is suppressed. 
The net effect is that star formation gradually shifts 
from the larger halos that populate overdense regions to  
lower density environments. 
Thus, a lack of cold gas due to gravitational heating in dense regions 
may provide a physical explanation of cosmic downsizing and one of its 
manifestations is observed as the colour-density relation.

Summarising, our results are as follows.

\begin{itemize}
\item[1)]
The fraction of red, early-type, low SFR galaxies is  higher
in superclusters of filament-type  than in superclusters of spider-type,
higher in superclusters than among  galaxies in low-density global
environment, and  higher
in all environments in rich  groups than in poor groups.
\item[2)]
Groups of equal richness host galaxies with higher stellar masses, 
and they have a larger fraction of early-type and red galaxies, 
and a higher fraction of low SFR galaxies if they are located 
in superclusters of filament morphology.
\item[3)]
In spider-type superclusters the red, high SFR galaxies have higher
stellar masses and lower values of environmental densities 
around them than in filament-type superclusters.
\item[4)]
The peculiar velocities of the main galaxies in groups in 
superclusters of filament morphology are higher than in
groups in superclusters of spider morphology.
In filament-type superclusters groups with higher peculiar velocities of 
their main galaxies  are located in higher
density environments than those with low peculiar velocities.
\item[5)] 
There are significant differences between galaxy populations 
of the individual richest superclusters.
\end{itemize}

These results suggest that both local (group) and 
global (supercluster) environments, and
inner structure and detailed density distribution in
superclusters are important factors affecting the evolution of galaxies and
galaxy groups in them. Our study indicated several open questions that need future
studies. 
It is not clear how the processes that shape the properties of
galaxies in different environments are related to the supercluster
scale density field, and how the supercluster morphology is related
to the properties of galaxies and groups in them.
We plan to develop further the methods to quantify inner structure of superclusters,
in order to understand better the role of supercluster
environment in the formation and evolution of 
galaxies and galaxy groups in them. We also plan to study the inner 
structure of superclusters using
galactic filaments \citep{2013arXiv1308.2533T}.

\section*{Acknowledgments}

We thank the referee for invaluable comments and suggestions 
which helped to improve the paper.
We thank Enn Saar for very fruitful collaboration on studies of supercluster
morphology.
We are pleased to thank the SDSS Team for the publicly available data
releases.  
Funding for the Sloan Digital Sky Survey (SDSS) and SDSS-II has been
  provided by the Alfred P. Sloan Foundation, the Participating Institutions,
  the National Science Foundation, the U.S.  Department of Energy, the
  National Aeronautics and Space Administration, the Japanese Monbukagakusho,
  and the Max Planck Society, and the Higher Education Funding Council for
  England.  The SDSS Web site is \texttt{http://www.sdss.org/}.
  The SDSS is managed by the Astrophysical Research Consortium (ARC) for the
  Participating Institutions.  The Participating Institutions are the American
  Museum of Natural History, Astrophysical Institute Potsdam, University of
  Basel, University of Cambridge, Case Western Reserve University, The
  University of Chicago, Drexel University, Fermilab, the Institute for
  Advanced Study, the Japan Participation Group, The Johns Hopkins University,
  the Joint Institute for Nuclear Astrophysics, the Kavli Institute for
  Particle Astrophysics and Cosmology, the Korean Scientist Group, the Chinese
  Academy of Sciences (LAMOST), Los Alamos National Laboratory, the
  Max-Planck-Institute for Astronomy (MPIA), the Max-Planck-Institute for
  Astrophysics (MPA), New Mexico State University, Ohio State University,
  University of Pittsburgh, University of Portsmouth, Princeton University,
  the United States Naval Observatory, and the University of Washington.

The present study was supported by the Estonian Science Foundation
grant No. 8005, grants 9428, MJD272, PUT246,  
by the Estonian Ministry for Education and
Science research project SF0060067s08, and by the European Structural Funds
grant for the Centre of Excellence "Dark Matter in (Astro)particle Physics and
Cosmology" TK120. This work has also been supported by
ICRAnet through a professorship for Jaan Einasto.
H. Lietzen acknowledges financial support from the Spanish Ministry
of Economy and Competitiveness (MINECO) under the 2011 Severo
Ochoa Program MINECO SEV-2011-0187.

\bibliographystyle{aa}
\bibliography{sclgal.bib}

\begin{appendix}

\section{Supercluster data }
\label{sect:cldata}


In Table~\ref{tab:scldata} we present data on superclusters.

\begin{table*}[ht]
\caption{Supercluster data}
\begin{tabular}{rrrrrrrrr} 
\hline\hline  
(1)&(2)&(3)&(4)&(5)& (6)&(7)&(8)&(9)\\      
\hline 
 ID(long)   &ID &  $N_{\mathrm{gal}}$ & $d$&$L_{\mathrm{tot}}$  &
$V_3$  & $K_1/K_2$ & Type & ID(E01)\\
 &  & & $h^{-1}$ Mpc&$10^{10} h^{-2} L_{\sun}$& &&&\\

\hline
239+027+0091 &   1 & 1041& 264& 1809   &   2.5  &   0.41  & F & 162    \\
227+006+0078 &   7 & 1217& 233& 1675   &   3.0  &   0.56  & S & 154    \\
184+003+0077 &  19 & 2060& 231& 2919   &   9.0  &   0.33  & S & 111    \\
167+040+0078 &  24 &  580& 225&  751   &   2.0  &   0.86  & S & 95    \\
202-001+0084 &  27 & 3222& 256& 5163   &  14.5  &   0.26  & F & 126    \\
173+014+0082 &  54 & 1341& 241& 2064   &   5.5  &   0.56  & S & 111    \\
189+017+0071 &  89 &  515& 212&  610   &   2.0  &   1.98  & S & 271    \\
215+048+0071 &  92 &  527& 213&  690   &   2.5  &   0.40  & S & ---    \\
230+027+0070 &  99 & 2047& 215& 2874   &  10.0  &   0.26  & S & 158    \\
203+059+0072 & 143 &  668& 211&  753   &   2.0  &   1.27  & F & 133    \\
172+054+0071 & 211 & 1439& 207& 1618   &   7.0  &   0.36  & F & 109    \\
187+008+0090 & 214 &  735& 268& 1218   &   6.0  &   0.28  & S & 111    \\
197+039+0073 & 218 &  272& 215&  337   &   1.0  & $-20.02$  & S & 274    \\
207+026+0067 & 219 &  985& 187& 1007   &   4.0  &   0.61  & S & 138    \\
207+028+0077 & 220 &  603& 226&  768   &   4.0  &   0.47  & F & 138    \\
255+033+0086 & 233 &  474& 259&  790   &   4.0  &   0.96  & F & 167    \\
122+035+0084 & 298 &  246& 246&  345   &   1.0  &   4.27  & S & 75    \\
159+004+0069 & 319 &  245& 207&  296   &   2.0  &   0.97  & S & 91    \\
195+019+0064 & 344 &  264& 192&  290   &   1.0  &  $-0.14$  & S & 273    \\
216+016+0051 & 349 &  335& 159&  284   &   1.0  &   0.97  & S & 143    \\
227+007+0045 & 352 &  519& 135&  379   &   1.0  &   1.46  & S & 154    \\
232+029+0066 & 360 &  311& 196&  330   &   2.0  &  $-0.54$  & S & 158    \\
146+043+0072 & 491 &  199& 217&  256   &   1.0  &  $-0.15$  & S & ---    \\
168+002+0077 & 499 &  408& 228&  517   &   3.0  &   0.67  & F & 91    \\
176+055+0052 & 515 &  457& 155&  390   &   3.0  & $-11.10$  & S & 109    \\
208+046+0062 & 532 &  297& 189&  293   &   2.0  &   0.49  & F & ---    \\
214+002+0053 & 540 &  422& 163&  358   &   1.5  &   0.49  & S & ---    \\
223+016+0045 & 541 &  299& 135&  214   &   1.0  &  $-0.88$  & S & ---    \\
129+028+0079 & 721 &   94& 237&  145   &   1.0  &   0.22  & S & 76    \\
151+054+0047 & 782 &  652& 139&  465   &   3.0  &   0.38  & S & ---    \\
169+029+0046 & 793 &  211& 142&  156   &   1.0  &  $-1.24$  & S & 93    \\
176+015+0069 & 797 &  160& 205&  203   &   1.0  &   0.22  & S & ---    \\
240+053+0065 & 849 &  215& 194&  242   &   1.0  &  $-0.03$  & F & 162    \\
244+049+0057 & 850 &  135& 171&  128   &   1.0  &   0.63  & S & 162    \\
234+036+0065 & 865 &  161& 197&  170   &   2.0  &  $-1.68$  & S & 158    \\
246+014+0050 & 868 &  239& 153&  201   &   1.5  &   0.98  & S & ---    \\
219+024+0087 & 870 &   76& 261&  143   &   1.0  &  $-0.14$  & S & ---    \\
127+030+0051 &1104 &  127& 150&  110   &   1.0  &   0.68  & S & ---    \\
191+054+0085 &1192 &   79& 248&  134   &   1.0  &   1.41  & S & ---    \\
220+010+0051 &1238 &  133& 156&   97   &   1.0  &  $-0.28$  & F & ---    \\
226+028+0057 &1244 &  101& 174&   97   &   1.0  &   0.76  & S & 152    \\
223+018+0059 &1247 &  147& 176&  149   &   1.0  &   0.28  & F & ---    \\
\hline
\label{tab:scldata}  
\end{tabular}\\
\tablefoot{                                                                                 
The columns are:
1: ID of the supercluster AAA+BBB+ZZZZ, 
where AAA is R.A., +/-BBB is Dec. (in degrees), and ZZZZ is 1000$z$;
2: ID of the supercluster from the SDSS DR8 superclusters catalogue \citep{2012A&A...542A..36E};
3: the number of galaxies in the supercluster, $N_{\mbox{gal}}$;
4: the distance of the supercluster;
5: the total weighted luminosity of galaxies in the supercluster, $L_{\mbox{tot}}$;
6: the maximum value of the fourth Minkowski functional,
$V_{3,\mathrm{max}}$ (clumpiness), for the supercluster;
7: the ratio of the shapefinders $K_1/K_2$ (the shape parameter) for the supercluster;
8: Morphological type of a supercluster
9: ID(E01): the supercluster ID  in  the catalogue by \citet{2001AJ....122.2222E}. 
SCl~111 and SCl~126 -- members of the Sloan Great Wall, SCl~158 -- the Corona 
Borealis supercluster, SCl~138 -- the Bootes supercluster, SCl~109 -- the Ursa 
Major supercluster.
}
\end{table*}

\section{Minkowski functionals and shapefinders} 
\label{sect:MF}

For a given surface the four Minkowski functionals (from the first to the
fourth) are proportional to the enclosed volume $V$, the area of the surface
$S$, the integrated mean curvature $C$, and the integrated Gaussian curvature
$\chi$. 
Consider an
excursion set $F_{\phi_0}$ of a field $\phi(\mathbf{x})$ (the set
of all points where the density is higher than a given limit,
$\phi(\mathbf{x}\ge\phi_0$)). Then, the first
Minkowski functional is the volume of this region (the excursion set):
\begin{equation}
\label{mf0}
V_0(\phi_0)=\int_{F_{\phi_0}}\mathrm{d}^3x\;.
\end{equation}
The second Minkowski functional is proportional to the surface area
of the boundary $\delta F_\phi$ of the excursion set:
\begin{equation}
\label{mf1}
V_1(\phi_0)=\frac16\int_{\delta F_{\phi_0}}\mathrm{d}S(\mathbf{x})\;.
\end{equation}
The third Minkowski functional is proportional to the
integrated mean curvature $C$ of the boundary:
\begin{equation}
\label{mf2}
V_2(\phi_0)=\frac1{6\pi}\int_{\delta F_{\phi_0}}
    \left(\frac1{R_1(\mathbf{x})}+\frac1{R_2(\mathbf{x})}\right)\mathrm{d}S(\mathbf{x})\;,
\end{equation}
where $R_1(\mathbf{x})$ and $R_2(\mathbf{x})$ 
are the principal radii of curvature of the boundary.
The fourth Minkowski functional is proportional to the integrated
Gaussian curvature (the Euler characteristic) 
of the boundary:
\begin{equation}
\label{mf3}
V_3(\phi_0)=\frac1{4\pi}\int_{\delta F_{\phi_0}}
    \frac1{R_1(\mathbf{x})R_2(\mathbf{x})}\mathrm{d}S(\mathbf{x})\;.
\end{equation}
At high (low) densities this functional gives us the number of isolated 
clumps (void cavities) in the sample 
\citep{2002sgd..book.....M,saar06}:

\begin{equation}
\label{v3}
V_3=N_{\mbox{clumps}} + N_{\mbox{cavities}} - N_{\mbox{tunnels}}.
\end{equation}

\begin{figure}[ht]
\centering
\resizebox{0.445\textwidth}{!}{\includegraphics[angle=0]{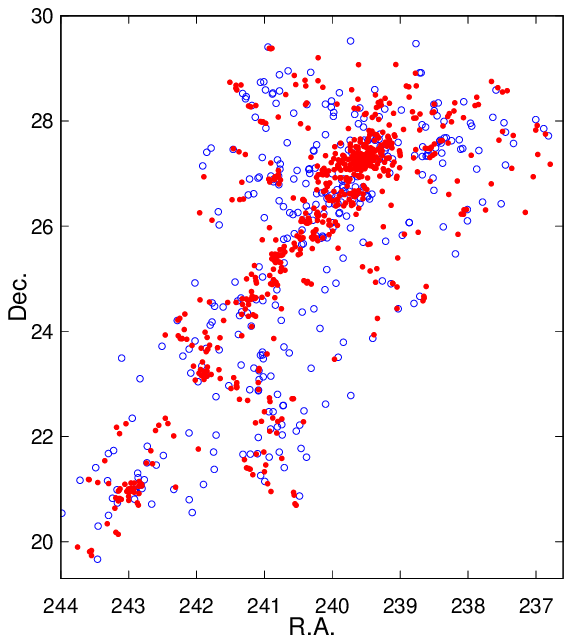}}
\resizebox{0.47\textwidth}{!}{\includegraphics[angle=0]{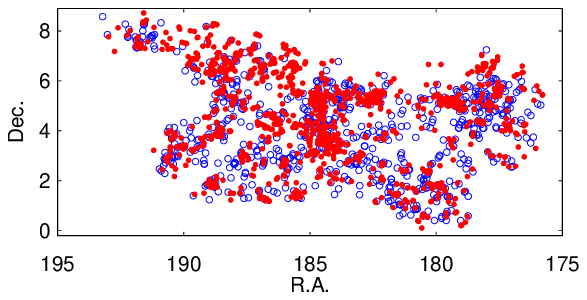}}
\caption{
Sky distribution of galaxies for the supercluster SCl~001
of filament morphology (upper panel), and for the supercluster SCl~019 
of spider morphology (lower panel). 
Red dots correspond to galaxies with low SFR, 
$\log \mathrm{SFR} < -0.5$,
and blue dots to galaxies with high SFR,
$\log \mathrm{SFR} \geq -0.5$. 
}
\label{fig:filspimorph}
\end{figure}

The first three functionals were used to calculate the shapefinders
$K_1$ (planarity) and $K_2$ (filamentarity)
\citep{sah98,sss04,saar09} and their ratio, the shape parameter 
$K_1$/$K_2$. 
The smaller the shape parameter, the more elongated a supercluster is.
We used morphological
information about superclusters, and their visual appearance
to
classify them as filaments and spiders. As an example, we present
in Fig.~\ref{fig:filspimorph}
the sky distribution of galaxies in one filament-type and one 
spider-type supercluster (SCL~001 and SCL~019, respectively). 

\section{Quantiles of galaxy parameters in different samples.} 
\label{sect:quantiles}

\begin{table}[ht]
\caption{Quantiles of galaxy properties in superclusters 
of filament and spider morphology, and in the field.}
\begin{tabular}{rrrr} 
\hline\hline  
(1)&(2)&(3)&(4) \\      
\hline 
Sample  & $1stQ$  & $\mathrm{median}$ & $3rdQ$  \\
\hline
          &   \multicolumn{3}{c}{$g - r$} \\       
Filaments &   0.670 &  0.769  & 0.805    \\      
Spiders   &   0.662 &  0.764  & 0.803    \\      
Field     &   0.591 &  0.722  & 0.787    \\      
          &   \multicolumn{3}{c}{$p_l$} \\       
Filaments &   0.088 &  0.500  & 0.897    \\      
Spiders   &   0.147 &  0.647  & 0.915    \\      
Field     &   0.196 &  0.763  & 0.931    \\      
          &   \multicolumn{3}{c}{$\log SFR$} \\       
Filaments &  -1.377 & -0.932  & 0.066    \\      
Spiders   &  -1.368 & -0.909  & 0.083    \\      
Field     &  -1.207 & -0.262  & 0.249    \\      
          &   \multicolumn{3}{c}{$\log M_\mathrm{s}$} \\       
Filaments &  10.390 & 10.590  & 10.840   \\       
Spiders   &  10.370 & 10.580  & 10.830   \\       
Field     &  10.300 & 10.530  & 10.770   \\       
\hline
                                                               
\label{tab:filspiisopara}                                                          
\end{tabular}\\
\tablefoot{                                                                                 
Columns are as follows:
1: Sample. 
2--4: Values of quantiles of galaxy properties 
(1st quantile, median, and 3rd quantile).
}
\end{table}

\begin{table}[ht]
\caption{Quantiles of the properties of high and low stellar mass
galaxies in superclusters 
of filament and spider morphology, and in the field.}
\begin{tabular}{rrrr} 
\hline\hline  
(1)&(2)&(3)&(4) \\      
\hline 
Sample   &  $1stQ$  & $\mathrm{median}$ & $3rdQ$  \\
\hline
$\log M_\mathrm{s} < 10.25$ &    \multicolumn{3}{c}{$g - r$} \\          
Filaments &  0.459& 0.528 & 0.606  \\
Spiders   &  0.465& 0.534 & 0.627  \\
Field     &  0.451& 0.512 & 0.575  \\
$\log M_\mathrm{s} \geq 10.25$ &     \multicolumn{3}{c}{$g - r$} \\      
Filaments &  0.723 & 0.778 & 0.809  \\ 
Spiders   &  0.715 & 0.776 & 0.807  \\ 
Field     &  0.668 & 0.755 & 0.797  \\ 
$\log M_\mathrm{s} < 10.25$ &    \multicolumn{3}{c}{$p_l$} \\  
Filaments &  0.714 & 0.889  & 0.938  \\ 
Spiders   &  0.770 & 0.898  & 0.946  \\ 
Field     &  0.806 & 0.905  & 0.947  \\ 
$\log M_\mathrm{s} \geq 10.25$ &  \multicolumn{3}{c}{$p_l$} \\ 
Filaments &   0.077 & 0.393  & 0.866  \\ 
Spiders   &   0.117 & 0.514  & 0.896  \\ 
Field     &   0.136 & 0.618  & 0.919  \\ 
$\log M_\mathrm{s} < 10.25$ &  \multicolumn{3}{c}{$\log SFR$} \\ 
Filaments &   -0.003 & 0.190  & 0.321  \\ 
Spiders   &   -0.017 & 0.174  & 0.316  \\ 
Field     &    0.045 & 0.197  & 0.330  \\ 
$\log M_\mathrm{s} \geq 10.25$ & \multicolumn{3}{c}{$\log SFR$} \\ 
Filaments &   -1.418 &-1.076  &-0.191  \\ 
Spiders   &   -1.422 &-1.064  &-0.171  \\ 
Field     &   -1.303 &-0.742  & 0.165  \\ 
\hline
\label{tab:filspiisohilosmass}                                                          
\end{tabular}\\
\tablefoot{                                                                                 
Columns are as follows:
1: Sample. 
2--4: Values of quantiles of galaxy properties 
(1st quantile, median, and 3rd quantile).
}
\end{table}

\begin{table}[ht]
\caption{Quantiles of the properties of red, low SFR (1, see {\it Notes}),
red, high SFR (2), and blue, high SFR (3)
galaxies in superclusters of filament and spider morphology.}
\begin{tabular}{rrrrr} 
\hline\hline  
(1)&(2)&(3)&(4)&(5) \\      
\hline 
Sample   & Subsample  & $1stQ$  & $\mathrm{median}$ & $3rdQ$  \\
\hline
          &&   \multicolumn{3}{c}{$D_8$} \\       
Filaments & $1  $ &  6.055  & 7.376 &   9.518  \\
          & $2 $ &  5.619  & 6.702 &   8.214  \\
          & $3 $ &  5.526  & 6.528 &   7.988  \\
        &          &        &         &          \\       
Spiders   & $1 $ &  5.772  & 7.316 &   9.381  \\
          & $2 $ &  5.576  & 6.618 &   8.324  \\
          & $3 $ &  5.488  & 6.514 &   8.392  \\
          & &  \multicolumn{3}{c}{$p_l$} \\       
Filaments & $1 $ &  0.062  & 0.213 &   0.648  \\ 
          & $2 $ &  0.232  & 0.807 &   0.951  \\ 
          & $3 $ &  0.684  & 0.896 &   0.945  \\ 
        &          &        &         &          \\       
Spiders   & $1 $ &  0.078  & 0.264 &   0.714  \\
          & $2 $ &  0.448  & 0.875 &   0.964  \\ 
          & $3 $ &  0.808  & 0.914 &   0.952  \\ 
          & &  \multicolumn{3}{c}{$\log M_\mathrm{s}$} \\       
Filaments & $1 $ & 10.50   &10.68  &  10.91   \\
          & $2 $ & 10.53   &10.70  &  10.92   \\
          & $3 $ & 10.12   &10.31  &  10.53   \\
          &      &        &         &          \\       
Spiders   & $1 $ & 10.47   &10.67  &  10.89   \\
          & $2 $ & 10.54   &10.74  &  10.95   \\
          & $3 $ & 10.10   &10.30  &  10.52   \\
\hline
\label{tab:filspihilosfr}                                                          
\end{tabular}\\
\tablefoot{                                                                                 
Columns are as follows:
1: Supercluster sample. 
2: Galaxy populations divided by colour and SFR. 1: red, low SFR galaxies with
$g - r \geq 0.7$, and $\log \mathrm{SFR} < -0.5$; 
2: red, high SFR galaxies
with $g - r \geq 0.7$, and $\log \mathrm{SFR} \geq -0.5$ 
and 3: blue, high SFR galaxies with 
$g - r < 0.7$, and $\log \mathrm{SFR} \geq -0.5$. 
3--5: Values of quantiles of galaxy properties 
(1st quantile, median, and 3rd quantile).
}
\end{table}

\begin{table}[ht]
\caption{Quantiles of the properties of 
galaxies in groups of different richness 
in superclusters of filament and spider morphology.}
\begin{tabular}{rrrrr} 
\hline\hline  
(1)&(2)&(3)&(4)&(5) \\      
\hline 
Sample   & Subsample  & $1stQ$  & $\mathrm{median}$ & $3rdQ$  \\
\hline
          &&   \multicolumn{3}{c}{$\log M_\mathrm{s}$} \\       
Filaments & $ 1 - 3 $ & 10.33  & 10.54  & 10.79  \\ 
          & $ 4 - 11$ & 10.41  & 10.61  & 10.86  \\ 
          & $\geq 12$ & 10.43  & 10.63  & 10.87  \\ 
        &          &        &         &          \\       
Spiders   & $ 1 - 3 $ & 10.29  & 10.53  & 10.79  \\ 
          & $ 4 - 11$ & 10.36  & 10.58  & 10.84  \\ 
          & $\geq 12$ & 10.41  & 10.61  & 10.85  \\ 
         & &   \multicolumn{3}{c}{$SFR$} \\       
Filaments & $ 1 - 3 $ &  -1.219 & -0.296 &   0.246 \\  
          & $ 4 - 11$ &  -1.355 & -0.927 &   0.075 \\  
          & $\geq 12$ &  -1.462 & -1.149 &  -0.428 \\  
        &          &        &         &          \\       
Spiders   & $ 1 - 3 $ &  -1.219 & -0.276 &   0.236 \\  
          & $ 4 - 11$ &  -1.309 & -0.767 &   0.136 \\  
          & $\geq 12$ &  -1.458 & -1.142 &  -0.348 \\  
         & &   \multicolumn{3}{c}{$g - r$} \\       
Filaments & $ 1 - 3 $ &  0.605 & 0.727  &  0.790  \\ 
          & $ 4 - 11$ &  0.670 & 0.771  &  0.803  \\ 
          & $\geq 12$ &  0.734 & 0.784  &  0.813  \\ 
        &          &        &         &          \\       
Spiders   & $ 1 - 3 $ &  0.595 & 0.722  &  0.786  \\ 
          & $ 4 - 11$ &  0.646 & 0.758  &  0.799  \\ 
          & $\geq 12$ &  0.726 & 0.781  &  0.811  \\ 
         & &   \multicolumn{3}{c}{$p_l$} \\       
Filaments & $ 1 - 3 $ &  0.118 & 0.693  &  0.921  \\ 
          & $ 4 - 11$ &  0.068 & 0.438  &  0.897  \\ 
          & $\geq 12$ &  0.082 & 0.373  &  0.842  \\ 
        &          &        &         &          \\       
Spiders   & $ 1 - 3 $ &  0.242 & 0.790  &  0.935  \\ 
          & $ 4 - 11$ &  0.129 & 0.645  &  0.918  \\ 
          & $\geq 12$ &  0.115 & 0.511  &  0.884  \\ 
\hline
\label{tab:filspinggr}                                                          
\end{tabular}\\
\tablefoot{                                                                                 
Columns are as follows:
1: Sample. 
2: Group richness. 
3--5: Values of quantiles of galaxy properties 
(1st quantile, median, and 3rd quantile).
}
\end{table}

\end{appendix}

\end{document}